\documentclass[a4paper,twocolumn,aps,floatfix,citeautoscript,superscriptaddress,prl,floatfix]{revtex4-1}
\usepackage{epsfig}
\usepackage{epstopdf}
\usepackage{graphicx}
\usepackage{amsfonts,amsmath,latexsym,amssymb}
\usepackage{bm}
\usepackage{color}
\usepackage{blindtext}

\begin{document} 

\newcommand{\sagc}[1]{\textsf{\color[rgb]{0,0.2,0.1}(SAG: #1)}}
\newcommand{\sag}[1]{{\color[rgb]{0,0.4,0.1} #1}}

\newcommand{\dto}{Dy$_2$Ti$_2$O$_7$}
\newcommand{\hto}{Ho$_2$Ti$_2$O$_7$}
\newcommand{\Dy}{DTO}
\newcommand{\Ho}{HTO}
\newcommand{\Tb}{Tb$_2$Ti$_2$O$_7$}
\newcommand{\Ir}{Pr$_2$Ir$_2$O$_7$}
\newcommand{\R}{R$_2$M$_2$O$_7$}

\title{Dynamics and thermodynamics of a topological transition in spin ice materials under strain }

\author{ L. Pili}
\affiliation{{ Instituto de F\'{\i}sica de L\'{\i}quidos y Sistemas Biol\'ogicos (IFLYSIB), UNLP-CONICET, La Plata, Argentina}}
\affiliation{Departamento de F\'{\i}sica, Facultad de Ciencias Exactas, Universidad Nacional de La Plata, La Plata, Argentina}

\author{ A. Steppke}
\affiliation{Max Planck Institute for Chemical Physics of Solids,  01187 Dresden, Germany}

\author{ M. E. Barber}
\affiliation{Max Planck Institute for Chemical Physics of Solids,  01187 Dresden, Germany}

\author{F. Jerzembeck} 
\affiliation{Max Planck Institute for Chemical Physics of Solids,  01187 Dresden, Germany}

\author{C. W. Hicks}
\affiliation{Max Planck Institute for Chemical Physics of Solids,  01187 Dresden, Germany}

\author{P.C. Guruciaga}\thanks{currently at: European Molecular Biology Laboratory (EMBL), 69117 Heidelberg, Germany}
\affiliation{Centro At\'omico Bariloche, Comisi\'on Nacional de Energ\'{\i}a At\'omica (CNEA), Consejo Nacional de Investigaciones Cient\'{\i}ficas y T\'ecnicas (CONICET), San Carlos de Bariloche, R\'{\i}o Negro, Argentina}

\author{D. Prabhakaran}
\affiliation{Department of Physics, Clarendon Laboratory, University of Oxford, Park Road, Oxford, OX1 3PU, UK}

\author{R. Moessner}
\affiliation{Max Planck Institute for the Physics of Complex Systems, 01187 Dresden, Germany}

\author{ A. P. Mackenzie}
\affiliation{Max Planck Institute for Chemical Physics of Solids,  01187 Dresden, Germany}
\affiliation{School of Physics and Astronomy, University of St Andrews, United Kingdom}

\author{ S. A. Grigera}
\email[e-mail: ]{sag@iflysib.unlp.edu.ar}
\affiliation{{ Instituto de F\'{\i}sica de L\'{\i}quidos y Sistemas Biol\'ogicos (IFLYSIB), UNLP-CONICET, La Plata, Argentina}}
\affiliation{Departamento de F\'{\i}sica, Facultad de Ciencias Exactas, Universidad Nacional de La Plata, La Plata, Argentina}

\author{ R. A. Borzi}
\email[e-mail: ]{borzi@fisica.unlp.edu.ar}
\affiliation{{ Instituto de F\'{\i}sica de L\'{\i}quidos y Sistemas Biol\'ogicos (IFLYSIB), UNLP-CONICET, La Plata, Argentina}}
\affiliation{Departamento de F\'{\i}sica, Facultad de Ciencias Exactas, Universidad Nacional de La Plata, La Plata, Argentina}

\date{\today}

 \begin{abstract}
\noindent {We study single crystals of \dto\ and \hto\ under magnetic field and stress applied along their [001] direction.
We find that many of the features that the emergent gauge field of spin ice confers to the macroscopic magnetic properties are preserved in spite of the finite temperature. The magnetisation vs. field shows an \textit{upward} convexity within a broad range of fields, while the static and dynamic susceptibilities present a peculiar peak. Following this feature for both compounds, we determine a single experimental transition curve: that for the Kasteleyn transition in three dimensions, proposed more than a decade ago. Additionally, we observe that compression up to $-0.8\%$ along [001] does not significantly change the thermodynamics. However, the dynamical response of \hto\ is quite sensitive to changes introduced in the ${\rm Ho}^{3+}$ environment.  Uniaxial compression can thus open up experimental access to equilibrium properties of spin ice at low temperatures.}
 \end{abstract}

\maketitle

\textit{Motivation.} Finding macroscopic evidence for spin liquids is notoriously difficult~\cite{balents2010spin,knolle2019field}.  Spin ice materials are among the best understood classical instances of such topological magnets~\cite{bramwell2001spin,diep2004}, incorporating some of the features that usually serve to characterise them: an emergent gauge field~\cite{isakov2004dipolar,Henley2010,fennell2007pinch,fennell2009magnetic}, algebraic spin correlations~\cite{isakov2004dipolar} and fractionalised (local, gapped) excitations, known as magnetic monopoles~\cite{Castelnovo2008,morris2009}. Like a paramagnet, ideal spin ice at the lowest temperatures exhibits no ordering transition, and a large effective residual entropy~\cite{bramwell2020history}. 

The physical magnetisation of spin ice is directly related to the emergent gauge flux, to which an external field thus couples directly. A magnetic field 
$\textbf{B}||$[001] can thus induce a topological phase transition, known as the three-dimensional Kasteleyn transition, where 
in ideal conditions the entropy reduces to zero at a finite value of $B/T$, while the magnetisation $\textbf{M}$  saturates with a divergent susceptibility~\cite{jaubert2008three,jaubert2009kasteleyn}. We emphasize that  it is not an interaction which drives this non-analytical behaviour. Like quantum correlations destabilise an ideal Bose gas towards a condensate~\cite{Powell_2008},  the exchange interactions drop out as an energy scale among the large number of degenerate ground states, so that the non-trivial magnetisation curve scales with $B/T$ at low $T\ll J$. 

Such a transition, where $M$ grows with \textit{upward} convexity as a function of $B$, seems a conspicuous feature to detect. However a thorough experimental investigation of this physics --proposed in 2008-- has been hindered by two competing requirements: low temperatures and thermodynamic equilibrium. Here we combine the study of two materials with different magnetic energy scales, \hto\ (\Ho) and \dto\ (\Dy), to expose the topological aspect of the transition.
We also observe that uniaxial compression up to $0.8\%$ does not appreciably change the thermodynamic properties. In contrast, we find that \Ho\ dynamics is sensitive to changes in the ${\rm Ho}^{3+}$ environment.  Uniaxial compression thus opens the possibility of experimentally studying equilibrium properties of spin ice below its usual freezing point.

\textit{Background.} \dto\ and \hto\ are the canonical spin ice materials.  The rare earth magnetic moments, or \textit{spins}, are of Ising nature.    
They point radially towards or away from the centres of the tetrahedra at whose corners they sit (inset to Fig.~\ref{acChi}a)), forming a pyrochlore lattice. The effective nearest neighbour interaction, composed of exchange and dipolar terms, is ferromagnetic with a $J_{\rm eff} \approx 1.1~{\rm K}$ for \Dy\ and $\approx 1.8~{\rm K}$ for \Ho. This energy defines the nearest neighbours model (NN); within it,
the minimum energy value for any given tetrahedron is achieved for the six \textit{spin ice states}, with two spins pointing in and two out of it (empty tetrahedra in Fig.~\ref{acChi}a)), a degeneracy linked to the eponymous residual entropy of spin ice~\cite{bramwell2020history}. In turn, this local rule is what makes possible to interpret $\textbf{M}$ as a divergence free gauge field.
The breaking of the spin balance in a tetrahedron, with ``three-in/one-out" or ``one-in/three-out" configurations, can be viewed as a local excitation or \textit{monopole}\footnote{``All-in'' and ``all-out'' excitations are also possible, but they are so energetically costly that they do not need to be considered here.} (coloured spheres in Fig.~\ref{acChi}a)), which are crucial to understanding the materials' properties. The low temperature thermodynamics can be described as a fluid of magnetic monopoles~\cite{Castelnovo2008,morris2009}, while the dynamics of the magnetisation in canonical spin ices, in particular the freezing around $\approx 0.6~{\rm K}$~\cite{matsuhira2000low,snyder2004low,clancy2009revisiting}, is linked to their abundance and mobility~\cite{bramwell2020history,Jaubert2009nat}.
The strong long-range dipolar interactions between spins (the magnetic moments in \Dy\ and \Ho\ are $\approx 10\mu_B$) can be approximately taken into account by two terms: the energy needed to create monopoles from the spin-ice manifold, and a \textit{magnetic} Coulomb interaction between them. This is the basis of the ``dumbbell model"  ~\cite{Castelnovo2008}.

At low temperatures, a magnetic field $\textbf{B}||$[001] polarises all spins, selecting a single saturated configuration from the vast spin ice manifold. If $T \ll J_{\rm eff}$, the density of monopoles ($\rho$) is exponentially small at any field. It follows that on decreasing $B$, the only mechanism that decreases $M$ and increases the entropy for this orientation is the introduction of extended excitations in the form of strings of reversed spins spanning the crystal (red arrows in Fig.~\ref{acChi}a)). Topologically, this is similar to the entry of field lines in a type II superconductor. Crucially, the topological change here leads to an asymmetrical susceptibility that diverges below the critical field $B_K$, but vanishes above it~\cite{jaubert2008three}, with $B_K/T = {\rm const.}$ for the NN model ~\cite{baez20163d}.  
At higher temperature, the presence of magnetic monopoles should allow for shorter, non-spanning strings, which alter the nature of the transition.
Some evidence of this has been found by comparing the experimental magnetisation for a single temperature with simulations (in DTO~\cite{jaubert2008three} and HTO~\cite{jaubert2009kasteleyn}). Magnetisation was combined with neutron scattering to identify a region in the $T - B$ plane where string excitations abound ~\cite{morris2009}, and implicit indications of topological sector fluctuations have been found~\cite{jaubert2013topological}. Here we will uncover systematic departures from pure paramagnetic behaviour, report the experimental phase diagram for the three-dimensional Kasteleyn transition, and  study the dynamics under field and strain in spin-ice.

\begin{figure}[ht!]
\includegraphics[width=1.00\columnwidth]{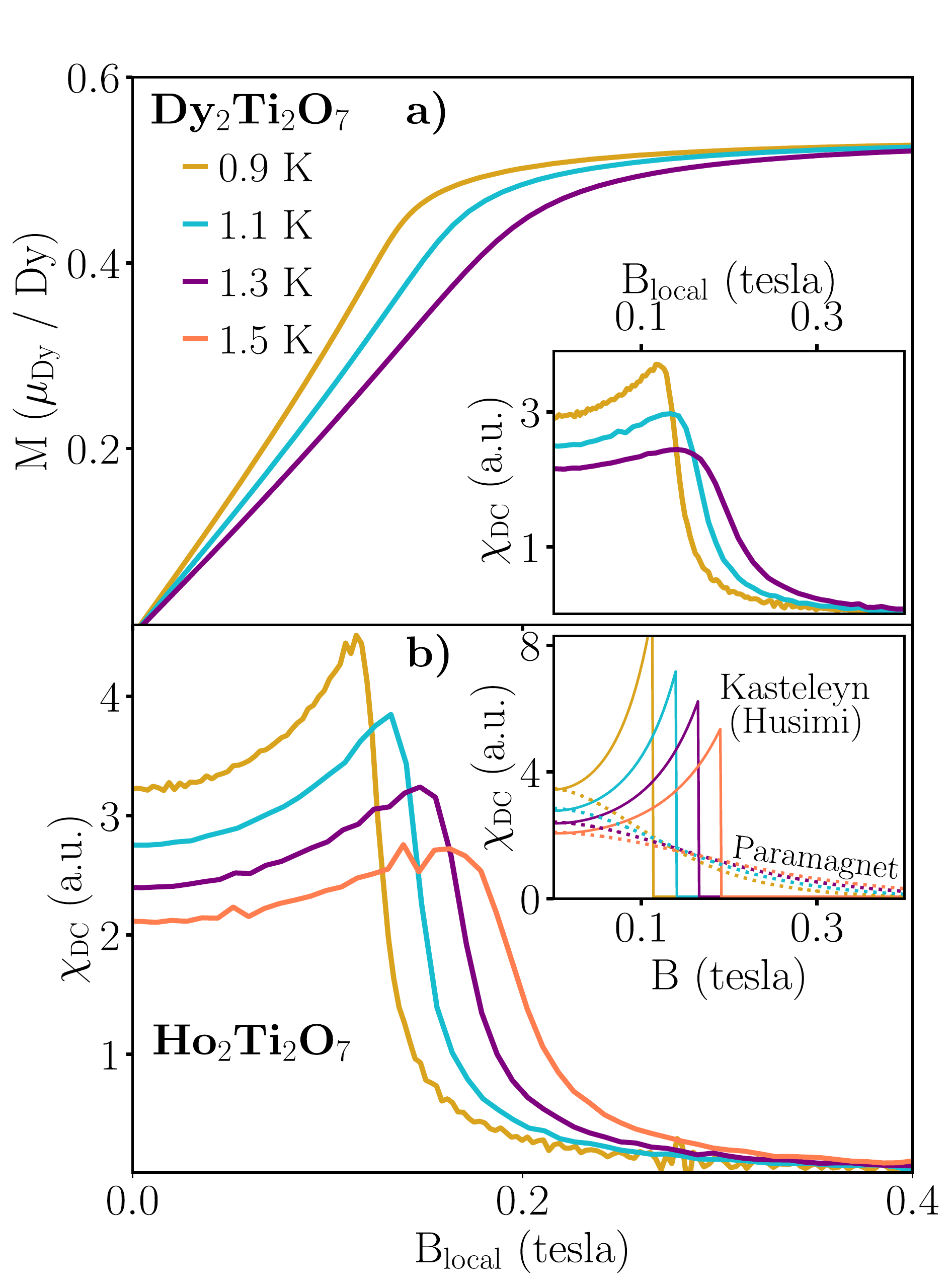}
	\caption{ a) Magnetisation ($M$) and static susceptibility ($\chi_{\text{\tiny{DC}}} \equiv dM/dB_{local}$, inset) for \Dy\ as a function of internal magnetic field along [001], $B_{\rm local}$, at different temperatures. 	b) $\chi_{\text{\tiny{DC}}}$ for \Ho\ as a function of internal magnetic field.  Inset: Ideal paramagnetic behaviour (dotted lines) and Husimi tree results for the NN model without monopole excitations. The shape and behaviour of the $\chi_{\text{\tiny{DC}}}$ peaks support a Kasteleyn transition broadened by thermal effects.}
 
	\label{M100}
\end{figure}

\begin{figure}[ht!]
	\includegraphics[width=1.00\columnwidth]{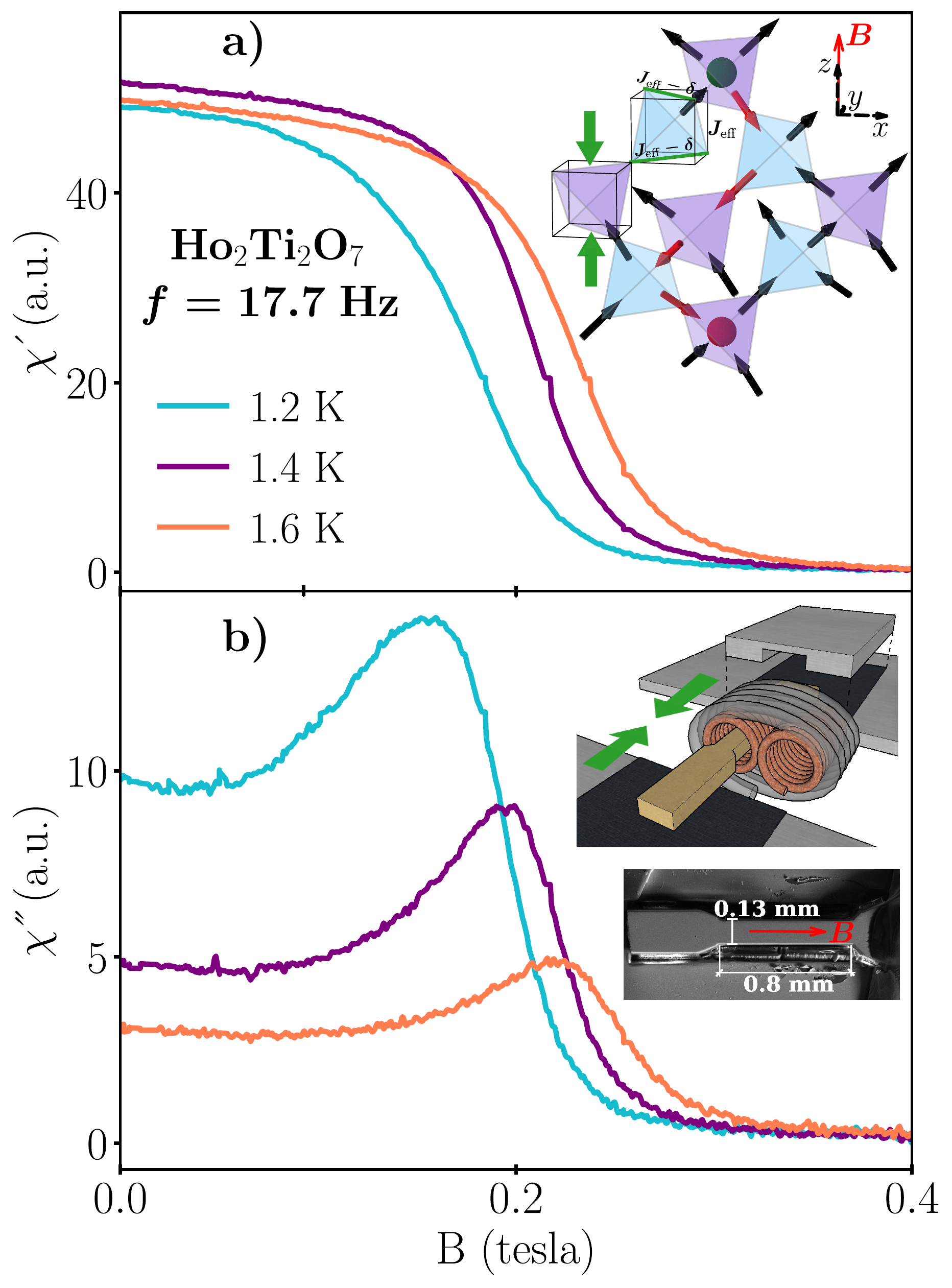}
	\caption{Real ($\chi'$) and Imaginary  ($\chi''$) parts of the ac-susceptibility in unstrained \Ho\ as a function of $\textbf{B}||$[001] for different temperatures. The monotonic  $\chi'(B)$ contrasts with the peaks in $\chi''$ at fields near those observed in $\chi_{\text{\tiny{DC}}}$. \textit{Inset to a):} Pyrochlore spin lattice. A strong $\textbf{B}||$[001] polarises all spins (black arrows); in red a string excitation. Tetrahedra with two spins in and two out, are neutral; an imbalance of spins results in a positive or negative magnetic charge (blue and red spheres). The green arrows mark the direction of compression. \textit{Inset to b):} \Ho\ crystal, cut in an hourglass shape along [001] (bottom); susceptometer and strain device (top). The superconducting primary (painted grey) winds around the two mutually opposing secondary coils (orange). The ends of the sample were epoxied to mobile piezoelectric anvils and covered by metallic plates.}
	\label{acChi}
\end{figure}

\textit{Results and Discussion.} We begin by analysing the unstrained samples. The isothermal magnetisation $M$ for both compounds was studied at temperatures above the dynamical freezing regimes ($T>0.7 {\rm ~K}$). Figure \ref{M100}a) shows $M(T,B)$ along the [001] crystalline axis of a \Dy\ single crystal as a function of field.  In the limit of high $B$ the magnetisation tends to saturate near $0.577 \mu_{Dy}$ per Dy, the theoretical value~\cite{fukazawa2002magnetic}. 
The magnetisation of an ideal paramagnet increases as a downwardly convex function for all fields; instead, the magnetisation curve of \Dy\ resembles that of a metamagnet, with an \textit{upward} convexity over a broad range of intermediate $B$. This is more clearly seen in the derivative $dM/dB_{local} \equiv \chi_{\text{\tiny{DC}}}$, the static susceptibility (see inset). The evolution of $\chi_{\text{\tiny{DC}}}$ at low fields and high $T$ suggest a single characteristic energy scaling with $B$. However, rather than the monotonically decreasing $\chi_{\text{\tiny{DC}}}$ of paramagnets (inset to Fig.~\ref{M100}b)), in \Dy\ the susceptibility \textit{increases} with field, and vanishes after it peaks. As the temperature is lowered the peak narrows and becomes more asymmetric. Sharper peaks, with the same characteristic asymmetry and at matching values of $B/T$ are seen in \Ho\ (Fig. \ref{M100}b)).  Since \Ho\ has an effective exchange interaction almost twice that of \Dy\, this coincidence in $B/T$ is already a strong indication that this feature, unlike in a metamagnet, is not a proportional to the strength of pair interactions. 

The Kasteleyn transition is indeed characterised by such a marked asymmetry in $\chi_{\text{\tiny{DC}}}$, since string excitations only exist on one side of the transition. This is clearly seen in the inset to Fig. \ref{M100}b), which shows a Husimi tree calculation of $\chi_{\text{\tiny{DC}}}$ for the NN model without monopolar excitations ($T \ll J_{\rm eff}$)\cite{JaubertThesis}.  This condition is not fulfilled in our experiments, and while $\chi_{\text{\tiny{DC}}}$ is rounded off by finite-size string-excitations at both sides of the transition (not unlike the effect of finite size~\cite{baez20163d}), the curves retain the main signatures of a Kasteleyn transition. The higher value of $J_{\rm eff}/T$ for \Ho\, explains why narrower, taller, and more asymmetric peaks are observed for \Ho ~\footnote{Our simulations for the dipolar Hamiltonian at $T=0.9$ and $B=0$ give $\approx 2\%$ of tetrahedra occupied by monopoles for \Dy: an order of magnitude bigger than for \Ho\ in the same conditions.}. 

In order to analyse the in-field dynamics, and to study the system under uniaxial pressure, we performed ac-susceptibility measurements.
Usually, a measurement at low frequencies ($\approx 10$ Hz) would give a real part of the susceptibility $\chi'$ almost identical to the static $\chi_{\text{\tiny{DC}}}$, and a small imaginary part $\chi''$~\cite{snyder2004low}.  This is not the case here, even at the lowest frequencies we measured ($\approx 1$ Hz). Fig.~\ref{acChi}a) shows $\chi'(B)$ at 17.7 Hz, measured for a single crystal of \Ho; within the temperature range inspected and for both compounds, we observed no peak in these curves. Remarkably, a peak instead appears in $\chi''(B)$ (see Fig.~\ref{acChi}b)), at fields near those found in $\chi_{\text{\tiny{DC}}}$. At a given $T$ the field position of the peak increases slightly with decreasing frequency (see Supp. Info.). The trend is more marked for \Dy\ (with slower dynamics at these temperatures), and increases with decreasing $T$. This reflects the dynamical nature of the peak in $\chi''$.

\begin{figure}[ht!]
	\includegraphics[width=0.95\columnwidth]{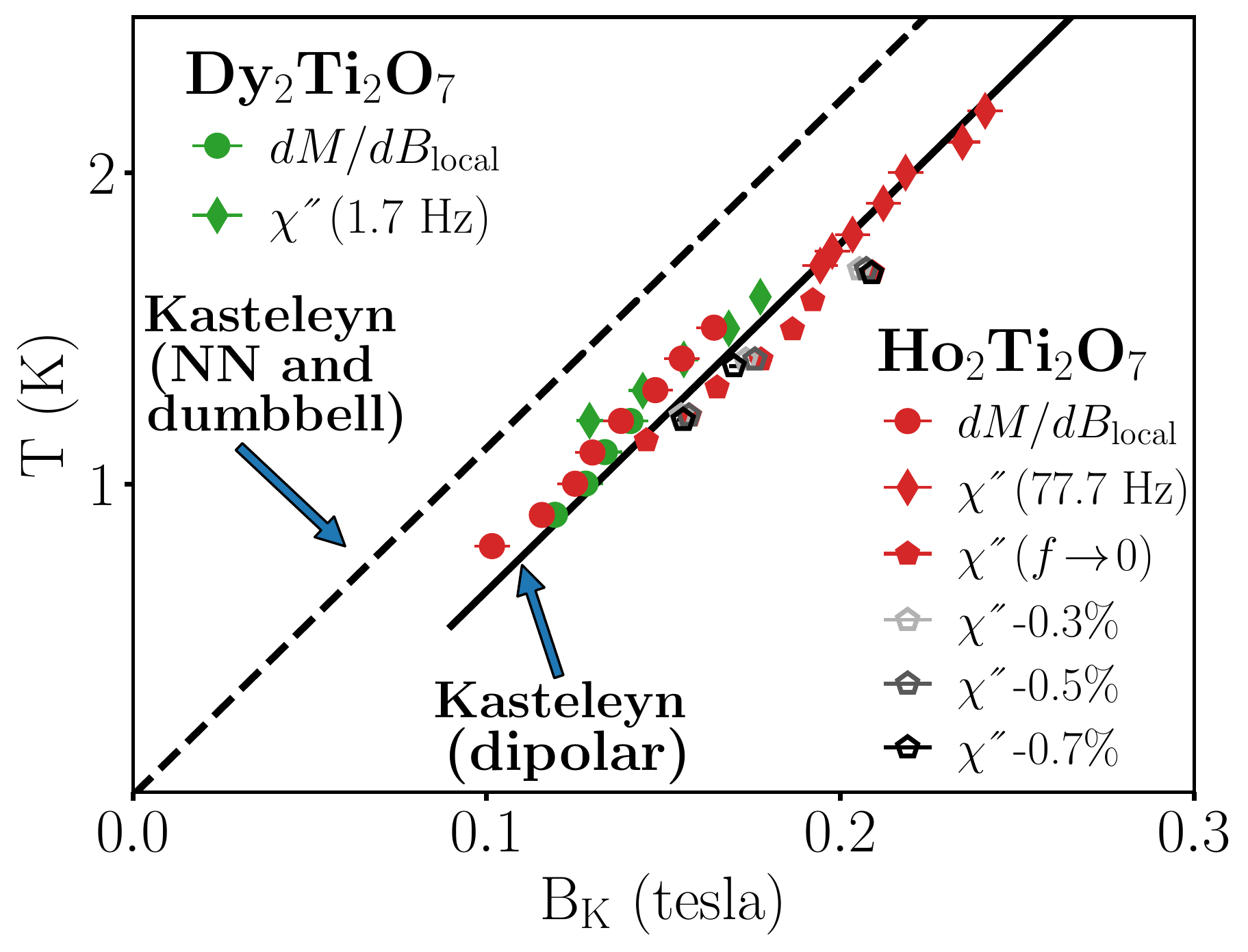}
	\caption{The experimental points, obtained for materials with different $J_{\rm eff}$ and diverse measurement techniques, determine a single line that separates two regions: a high $B/T$ region where the magnetisation is close to saturation and a low one where fluctuations proliferate.  The lines indicate the Kasteleyn transition curve for the NN and dumbbell models (dotted) and extrapolated from high $T$ for the dipolar model (solid). The open pentagons are the results for $\chi''$ ($f \to 0$) under strain.}
	\label{PhaseDiag}
\end{figure}

A common factor in both $\chi_{\text{\tiny{DC}}}$ and $\chi''$, for both materials, is to have a peak near $B/T \approx 0.1 ~{\rm tesla/K}$. This can be used to separate two regions in the field-temperature plane: a low field/high temperature region where extended fluctuations that lower the magnetisation proliferate; and a high field/low temperature one where fluctuations are scarce and the magnetisation is close to saturation.  Fig. \ref{PhaseDiag} shows a collection of data points associated with these maxima for both materials.
For ac-susceptibility the data is representative of different conditions, samples, and instruments, chosen to reflect information near the static limit. At low temperatures, the points collected for \Ho\ were obtained with our bespoke susceptometer at zero strain, after a linear extrapolation of the field maximum in $\chi''$ to the limit $f \rightarrow 0$ (see Supp. Info); those for \Dy\ were measured at $f=1.7$ Hz. The faster dynamics of \Ho\ at higher temperatures allowed us to use data obtained at $77$ Hz (with a bigger signal and less data treatment).

The data enable the construction of a curve that is, within errors, independent of the material, technique, and instrument used (Fig. \ref{PhaseDiag}). This independence allows us to associate this line of maxima with the phase diagram for the three-dimensional Kasteleyn transition for spin as if $T \ll J_{\rm eff}$~\cite{jaubert2008three,jaubert2009kasteleyn}. 
The dotted straight line in the figure indicates the predicted transition curve for a nearest-neighbour model \cite{jaubert2008three}.  The full line corresponds to this curve corrected by dipolar interactions~\cite{baez20163d}: it is a much better match to the experimental points.  At the mean field level the effect of dipolar interactions is to reduce the magnetic enthalpy for the creation of strings in the fully polarised phase, and thus to shift the transition along the field axis~\cite{baez20163d}.

The dumbbell approximation has successfully explained several dynamical~\cite{Jaubert2009nat,mostame2014tunable,Guruciaga2020,castelnovo2011DebyeH,bramwell2020history} and thermodynamical facts~\cite{Castelnovo2008,Guruciaga2014,morris2009,castelnovo2011DebyeH,bramwell2020history}.  Since the Kasteleyn transition takes place in the limit of no monopolar excitations, it should be identical in both the NN and the dumbbell model.  Higher-order dipolar corrections --similar for both materials-- account for the differences with the experimental results~\footnote{We have not tried to include other perturbation terms, such as second or third NN interactions~\cite{yavors2008dy,Borzi2016,henelius2016,samarakoon2020machine}.}.   
We find this remarkable, since thermodynamic data at low fields are expected to differ from the dumbbell model only when $T$ is of the order of a few hundred of mK~\cite{melko2001,pomaranski2013,Borzi2013,lin_kasteleyn,baez20163d}.

The change of dynamic regime marked by the maximum in $\chi''(T,B)$ is rooted in several facts. The depletion of magnetic charges is particularly important for $\textbf{B}||$[001], and quite abrupt near the transition~\cite{Guruciaga2020}. This lowers the number of spin flips at low temperatures, and reduces the screening~\cite{bramwell2020history}. Additionally, by polarising the magnetic moments, the static magnetic field changes the background in which the magnetic monopoles move. All this slows down the magnetic response of the system to an external field, reducing the blocking temperature~\cite{Guruciaga2020}. The peak in $\chi_{\text{\tiny{DC}}}$ observed in Fig.~\ref{M100} on decreasing field at low temperatures reflects the abrupt creation and/or marked development of long strings of inverted spins, driven by entropic forces. The absence of a peak in $\chi'(T,B)$ reflects the inability of these extended objects to oscillate in phase with the field even at frequencies as low as $1{\rm Hz}$; they are instead seen in $\chi''(T,B)$.

Our bespoke ac-susceptibility setup allows us to apply compressive uniaxial stress along the field direction (insets to Fig.~\ref{acChi}b)). Theoretical work~\cite{Jaubert2010Multicriticality} predicts that $B_K$ should decrease for crystals under compression along [001], through the progressive imbalance of the two pairs of exchange interactions (inset to Fig.~\ref{acChi}a)). A recent study in \Ho~\cite{edberg2020effects} reports a small evolution of $J_{\rm eff}$ with uniaxial pressure. Accordingly, the changes we observe in the position of the transition are almost negligible for the compression strengths that were experimentally accessible to us. This is seen in Figure~\ref{PhaseDiag}, where the open pentagons correspond to the $\chi''(B)$ peaks, after taking the $f \to 0$ limit, for different induced strain values. 

\begin{figure}[ht!]
	\includegraphics[width=0.95\columnwidth]{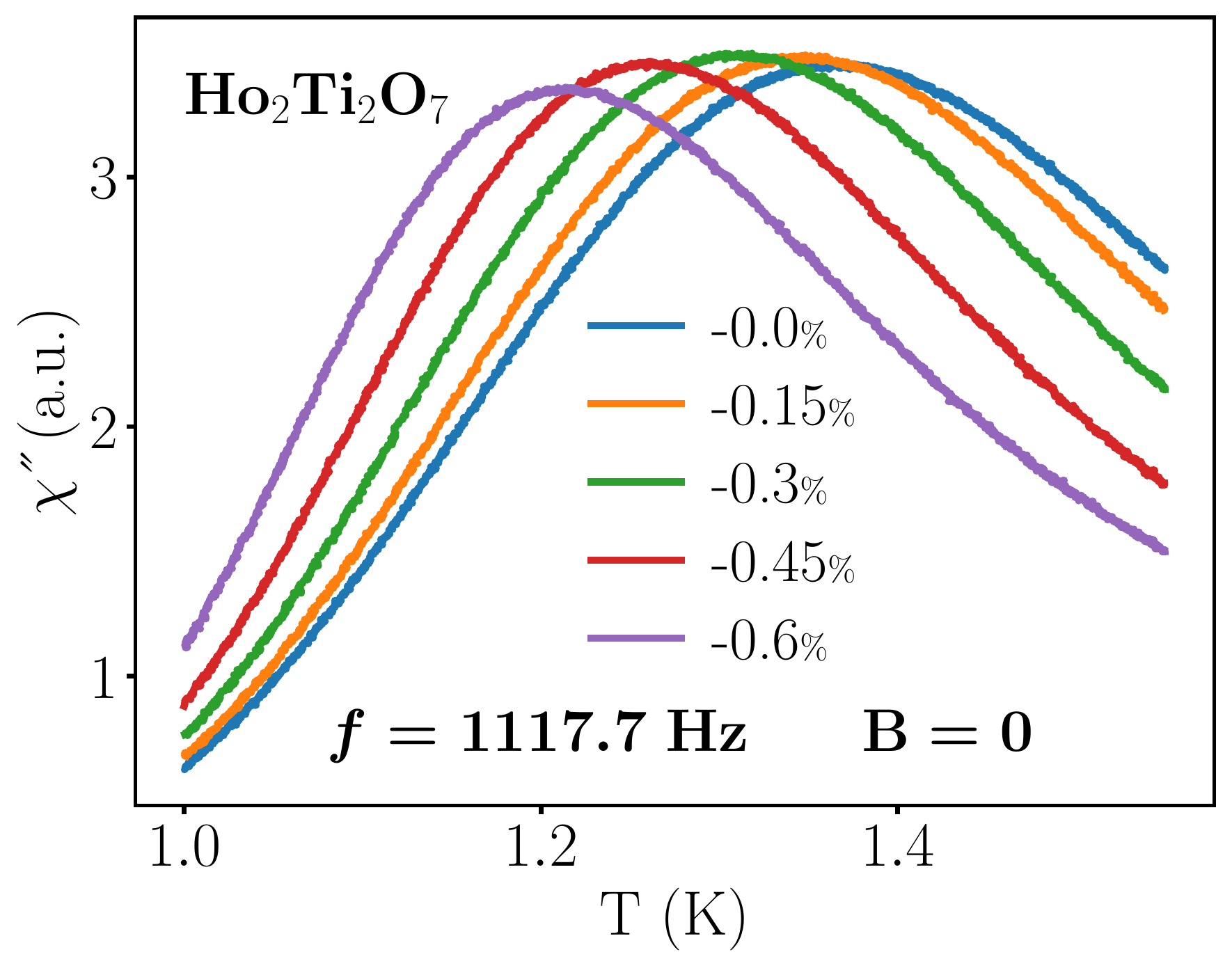}
	\caption{Imaginary part of the ac-susceptibility ($\chi''$) for \Ho\ at zero field, measured as a function of temperature at fixed frequency $f$ for different values of strain along [001]. The peak moves towards lower temperatures as compression increases, indicating a faster dynamics.}
	\label{dynamics}
\end{figure}

However, the effect of an applied stress on a crystal can go beyond thermodynamics. Altering the environment of a spin may enhance quantum tunnelling between its two Ising states and thus change its dynamical response \cite{tomasello2015single,tomasello2019correlated,bonville2011}. It has been theoretically proposed for \Tb, an Ising pyrochlore with physics less understood than spin ice~\cite{enjalran2004TBO}, that a tetragonal distortion can lead to faster dynamics~\cite{bonville2011}; this, together with the susceptibility of the non-Kramers \Ho\ ground state doublet to transverse fields ~\cite{tomasello2019correlated} might give the dynamics of \Ho\ a high sensitivity to uniaxial pressure. 
Figure \ref{dynamics} shows $\chi''_{\rm AC}$ at $f = 1117.7{\rm Hz}$ and $B=0$, as a function of temperature for different compressive strains. 
As seen there, the position of the peak is sensitive to the applied stress and moves markedly towards lower temperatures: at a given $T$ the response times shortens as the environment is distorted.
We believe that this observation ---combined with the lack of a measurable effect in the thermodynamic properties--- may open a whole avenue of research. Uniaxial stress in a material like \Ho, with a small exchange constant (dominated by the NN term in the dipolar interactions) could be the first route towards a canonical spin ice with faster dynamics which may reveal their long searched for ordered ground state\cite{melko2001,lin_kasteleyn,Borzi2016,henelius2016,Giblin_2018Pauling,samarakoon2020machine} and, perhaps, tunable quantum effects.

\textit{Conclusions.} By studying the magnetisation process under applied field  of two different spin-ice materials we have identified what could be called {\em topological metamagnetism}, i.e. a growth of the susceptibility with field resulting from the topological nature of the magnetisation, and independent of the interaction strength.  We constructed the phase diagram of its associated Kasteleyn transition and  showed that uniaxial stress applied to \hto\ leads to faster dynamics with canonical spin ice thermodynamic behaviour.

This work was carried out within the framework of a Max-Planck independent research group on strongly correlated systems. We acknowledge financial 
support from the Deutsche Forschungsgemeinschaft through SFB 1143 (project-id 247310070) and cluster of excellence ct.qmat (EXC 2147, project-id 390858490), EPSRC (EP/T028637/1), ShanghaiTech University, Agencia Nacional de Promoci\'on Cient\'\i fica y Tecnol\'ogica through PICT 2017-2347, and Consejo Nacional de Investigaciones Cient\'\i ficas y T\'ecnicas through PIP 0446.

\section{Supplementary Information}

\subsection{Materials and Experimental Details}\label{App:recast}

\textbf{Samples}. The \Dy\ and \Ho\ samples used in this paper were grown in St Andrews and Oxford using floating zone furnaces. For the strain measurements we used a \Ho\ single crystal, of approximate dimensions  $1.9~{\rm mm} \times 0.34~{\rm mm}$ and a thickness of $0.13~{\rm mm}$, cut along its principal axes. It was then etched using a Xenon Plasma Focused Ion Beam (FIB) into an hour-glass shape (see insets to Fig.~2 in the main text). This shape allowed to achieve high homogeneous compression in the middle section of the sample were the actual measurement took place. The  samples used for the other  measurements were two single crystals of Ho$_2$Ti$_2$O$_7$ and Dy$_2$Ti$_2$O$_7$, with approximate dimensions ($2.8 \times 0.75 \times 0.55$) ${\rm mm}^3$  and ($4.55 \times 0.71 \times 0.66$) ${\rm mm}^3$, respectively. In all samples, the susceptibility was measured with the longer side of the sample along the magnetic field direction in order to minimise demagnetisation effects.

\begin{figure}[ht!]
	\includegraphics[width=0.95\columnwidth]{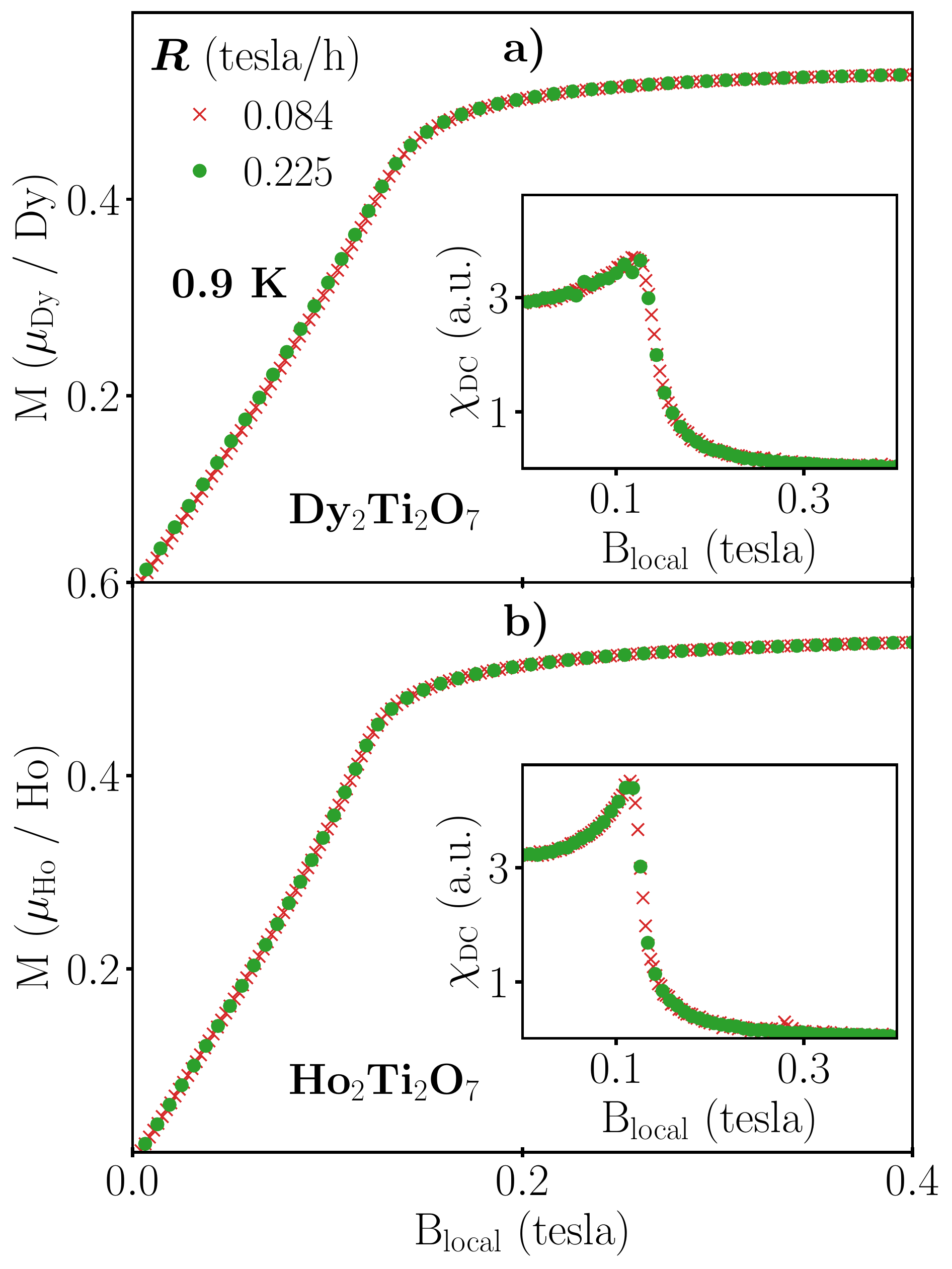}
	\caption{Magnetisation, $M$ and Static susceptibility, $\chi_{\text{\tiny{DC}}}$ (inset) at $0.9 {\rm K}$ for \Dy\ (panel a)) and \Ho\ (panel b)) as a function of internal magnetic field along [001], $B_{\rm local}$. We have used two field sweep rates $R \equiv dB/dt$ differing by a factor of $\approx 3$, obtaining the same results. This suggests that the measurements approach the static limit. It is easy to see that the susceptibility peak marking the Kasteleyn transition is much better defined and more asymmetric for \Ho, where the spin ice condition ``two spin in/two out'' at a given temperature is fulfilled for a much bigger fraction of tetrahedra.}
	\label{supp_rates}
\end{figure}

\textbf{Measurements.} The experiments were performed at two different locations, with several instruments and different samples. The magnetisation measurements were carried out using a commercial Quantum Design MPMS.  The standard field sweep rate used was $R \equiv dB/dt = 0.225~{\rm tesla/h}$. In order to check that these measurements were a good approximation to the static limit, we repeated the curves at low temperature (here, $0.9~{\rm K}$) at two different field-sweep rates, for both compounds. Fig.~\ref{supp_rates} shows the magnetisation and $\chi_{DC} \equiv dM/dB_{local}$ for \Dy\ (panel a)) and \Ho\ (panel b)). There is no evident field rate dependence in any of the compounds, in spite of $R$ varying by a factor near $3$.

We used three different probes for the ac-susceptibility measurements. The first part of the measurements were carried out with a bespoke probe in a single-shot $^3$He cryostat. The susceptometer consisted of a pair of counterwound pickup coils, each with approximately 1200 turns of 60 $\mu{\rm m}$ diameter copper wire. In order to guarantee a good thermal contact the probe was immersed in the He$^3$ chamber. The ac-field used was $1.75\times10^{-4}{\rm ~tesla}$.
 The second part was performed in a commercial Quantum-Design PPMS. Finally the susceptibility under applied strain (including compression free) was measured in a dilution refrigerator; in order to work at temperatures above $1.1{\rm K}$, the normal regime of operation was changed by partially decoupling the probe from the mixing chamber.
 
In this case, the extremely delicate pickup coils consisted of two counterwound secondary coils ($\approx 200$ turns each) of 11 $\mu{\rm m}$ diameter copper wire; the primary coil was wound around them with 20 turns of 60 $\mu{\rm m}$ superconducting wire (see Fig.2b) in the main text).  The excitation ac-field used was $\approx 8\times10^{-4}{\rm~tesla}$.

\begin{figure}[ht!]
	\includegraphics[width=1.00\columnwidth]{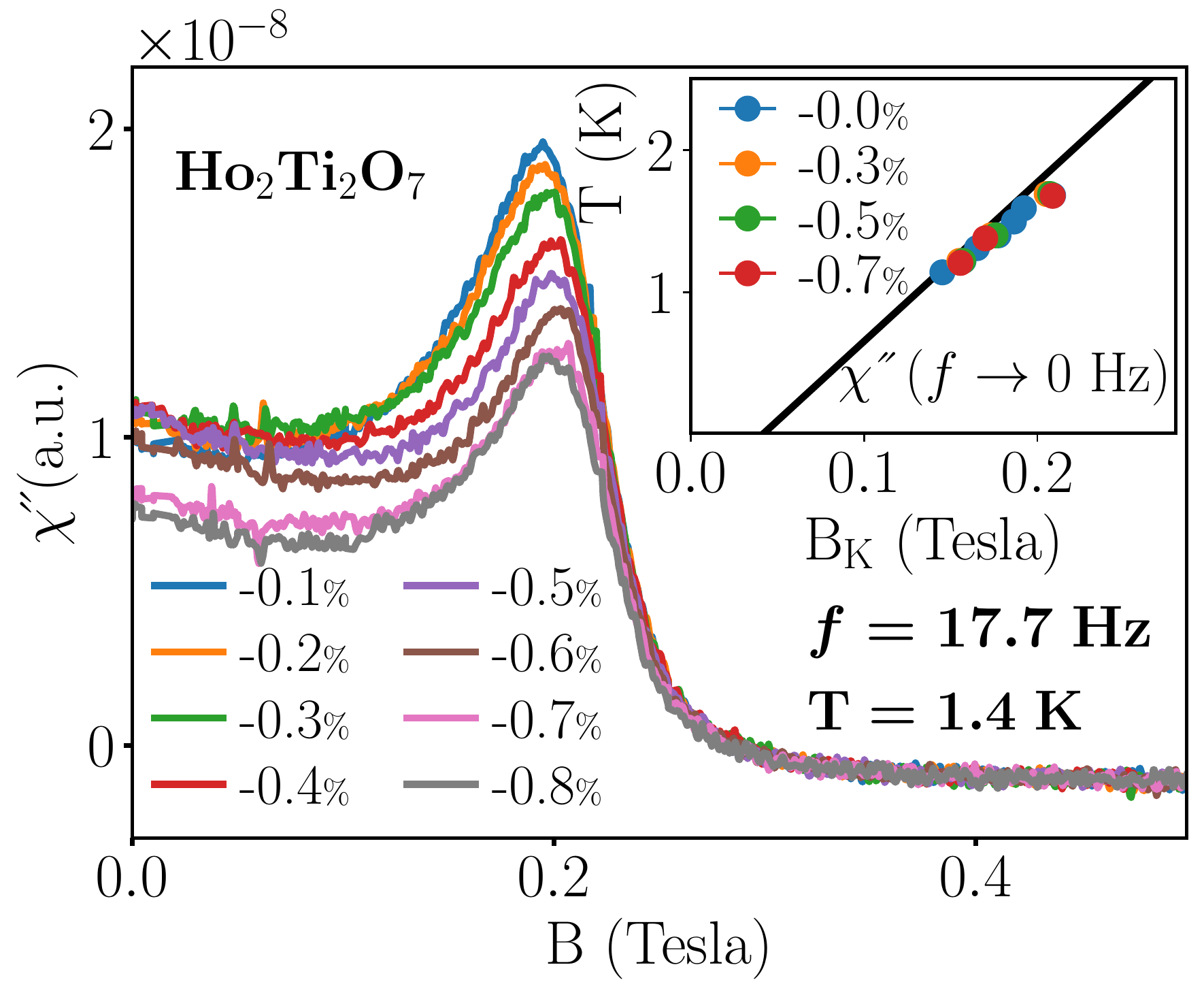}
	\caption{{\em Effects of compression.}  Imaginary part of the susceptibility  $\chi''_{\rm AC}$ vs field along [001] for different values of applied stress parallel to $\textbf{B}$. 
	In all cases the temperature is fixed at 1.4 K, and the frequency at $f=17.7~{\rm Hz}$. The inset shows the phase diagram under compressing strain for $f \to 0$.  The solid line indicates the predicted Kasteleyn transition in the presence of dipolar interactions under no applied stress.}
	\label{compress}
\end{figure}

The strain was applied using a piezoelectric-based uniaxial pressure cell with integrated force and displacement sensors developed at the Max Planck Institute for Chemical Physics of Solids in Dresden, Germany~\cite{hicks2014piezoelectric,barber2019piezoelectric}.

\textbf{Local vs. applied field.} The magnetic field $B$ was corrected taking into account the demagnetising factor $D$ in order to obtain the internal magnetic field $B_{\rm local}$ at the peak of the static or dynamic susceptibility.  $D$ was estimated with standard methods, assuming the samples were perfect rectangular prisms~\cite{aharoni1998demagnetizing}. In the case of ac-susceptibility, the magnetization at the peak of its imaginary part was taken from the MPMS measurements. We also estimated $B_{local}$ with the aid of Monte Carlo simulations; the difference between both procedures was used as a way to estimate the error in the peaks location. Due to the sample shape, the magnitude of the distortions, and the smallness of the demagnetisation factor, we did not consider changes in $D$ when straining the crystals.

\subsection{Imaginary part of the susceptibility under compression, and extrapolation to $f \to 0$}

Figure \ref{compress} shows $\chi''$ for \Ho\ as a function of $\textbf{B}||$[001] and $T=1.4{\rm K}$; different values of strain were induced along [001], between $-0.1\%$ and $-0.8\%$ (the minus sign here indicates compression).  At fixed $f=17.7~{\rm Hz}$ the effect of compression is very modest: it slightly \textit{increases} the field of the peak maximum. However, as explained below, extrapolating the data to $f \to 0$ this tendency is almost cancelled, indicating that the main effect of the uniaxial compressive stress is to accelerate the dynamics.

\begin{figure}[ht!]
\includegraphics[width=1.00\columnwidth]{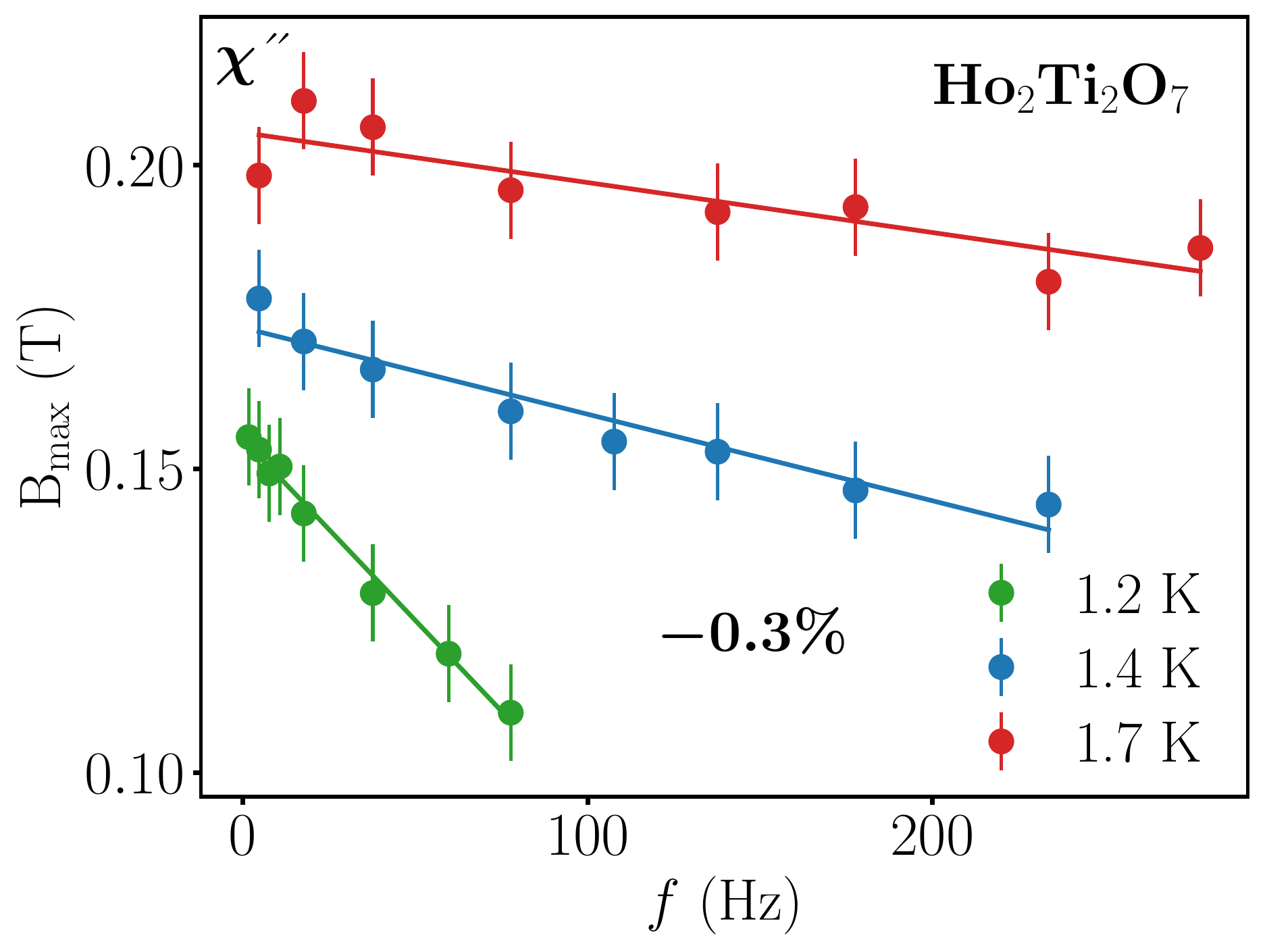}
	\caption{\textit{Extrapolation to $f \to 0$ for \Ho}. Each point represents the field maximum $B_{\rm max}$ in the imaginary part of the ac-susceptibility $\chi''(T,B)$ at a given temperature $T$ and frequency $f$, for a fixed strain of $-0.3\%$. $B_{max}$ increases with frequency, a trend that is more pronounced at low temperature. The points in the phase diagram corresponding to $f \to 0$ where taken after a linear extrapolation, as illustrated here.}
	\label{freq}
\end{figure}

As mentioned in the main text, the peak at the field $B_{\rm max}$ observed in the imaginary part of the susceptibility $\chi''$ not only depends on $T$, but also shows a smooth dependence with the frequency $f$. As illustrated in Fig.~\ref{freq}, this dependence is stronger at lower temperatures (i.e. greater relaxation times) and very small at high temperatures. One way of approach to the three dimensional Kasteleyn transition phase diagram is through the slow dynamics related with the creation or growth of sample spanning strings of inverted spins. In the main text we observed that the peak in $\chi''$ in the small frequency limit coincides with the phase transition peaks observed for both \Dy\ and \Ho\ in static measurements. Fig.~\ref{freq} illustrates how the limit $f \to 0$ was taken.

\bibliography{main}

\end{document}